\documentclass[a4paper]{article}
\def\x{{\mathbf x}}

\def\Y{{\mathbf Y}}
\def\X{{\mathbf X}}
\def\R{{\mathbf R}}
\def\S{{\mathbf S}}
\def\Z{{\mathbf Z}}
\def\P{{\mathbf P}}

\def\L{{\mathbf L}}
\usepackage{amsmath}
\newcommand{\comment}[1]{}
\usepackage{INTERSPEECH2019}

\title{Speaker embeddings by modeling channel-wise correlations}
\name{Themos Stafylakis$^{1}$, Johan Rohdin$^{1,2}$, Luk\'a\v{s} Burget$^2$}
\address{
$^1$Omilia - Conversational Intelligence, Athens, Greece\\
$^2$Brno University of Technology, Faculty of Information Technology, Speech@FIT, Czechia
\email{\{tstafylakis\}@omilia.com, \{rohdin,burget\}@fit.vutbr.cz}
}

\begin{document}

\maketitle
\begin{abstract}
Speaker embeddings extracted with deep 2D convolutional neural networks are typically modeled as projections of first and second order statistics of channel-frequency pairs onto a linear layer, using either average or attentive pooling along the time axis. In this paper we examine an alternative pooling method, where pairwise correlations between channels for given frequencies are used as statistics. The method is inspired by style-transfer methods in computer vision, where the style of an image, modeled by the matrix of channel-wise correlations, is transferred to another image, in order to produce a new image having the style of the first and the content of the second. By drawing analogies between image style and speaker characteristics, and between image content and phonetic sequence, we explore the use of such channel-wise correlations features to train a ResNet architecture in an end-to-end fashion. Our experiments on VoxCeleb demonstrate the effectiveness of the proposed pooling method in speaker recognition. 

\end{abstract}

\section{Introduction}
Extracting low-dimensional representations from speech utterances that characterize the speaker is a challenging tasks with numerous application. Apart from speaker recognition, speaker representations are used in automatic speech recognition (e.g. for speaker adaptation), in speaker diarization and separation, and in speech synthesis, for mimicking a target speaker using text-to-speech or voice conversion. 

Until recently, i-vectors were the most commonly used means for representing speaker characteristics \cite{kenny2007joint,dehak2011front}. With the advent of deep learning, neural architectures such as time-delay neural networks (TDNNs \cite{snyder2018x}), Long Short-Term Memory networks (LSTMs), 2D convolutional neural networks (such as ResNets and its extensions \cite{zhou2020resnext}) and Transformers have emerged, surpassing the performance of i-vectors in most applications and speaker recognition settings.

In order to obtain fixed-size representation vectors from utterances of variable duration and arbitrary word sequences (i.e. text-independent), a pooling method is required. The most common pooling method is a simple aggregation over the time axis, where the mean vector is concatenated with standard deviation (known as statistics pooling layer). The speaker embedding (a vector of few hundred dimensions) is obtained by projecting this pooled representation (a vector of few thousand dimensions) onto a linear layer or a shallow feed-forward neural network. Other pooling approaches have been proposed, such as single and multi-head attention, as well as the use of higher-order statistics \cite{okabe2018attentive,india2019self,you2019multi}.

In this paper we present an alternative way of modeling speaker information, that is largely inspired by style-transfer \cite{gatys2016image}. This seminal algorithm in computer vision is capable of producing new images that combine the content of an image with the style of other images or artworks. The style of an image is modeled by the correlations between the channels of the convolutional neural network. By considering styles as broadly analogous to speakers, we experiment with modeling speakers by channel-wise correlation. This is in contrast to averaging along the time dimension of each channel frequency pair of the output tensor that is used in statistics pooling. The other sources of variability that may also be considered as style of the recording, such as noise, channel, and emotion, can be suppressed by training the model end-to-end using supervised learning with speakers as targets. The resulting embeddings are highly speaker-discriminative, yielding state-of-the-art results on VoxCeleb using {\em plain} cosine similarity scoring.

The rest of the paper is organized as follows. In Sect. \ref{Sect:RW} we briefly review some of the related work, such as pooling methods and style-transfer. In Sect. \ref{Sect:Proposed} we introduce the proposed method, and provide all the implementation and architectural details. In Sect. \ref{Sect:Exps}, we present our experiments on a standard speaker recognition benchmark (VoxCeleb). Finally, in Sect. \ref{Sect:Conc} we provide conclusions and future work directions.

\section{Related work}
\label{Sect:RW}
\subsection{Style-transfer in computer vision and voice conversion}
A few years ago, the seminal work of L.A. Gatys {\it et al.} \cite{gatys2016image} in neural style-transfer gained the attention of the computer vision community and the industry, and received wide media coverage. Since then, several works extended their method, some of which can be found in \cite{jing2019neural}. A theoretical interpretation of the way style and image texture is encoded in the channel-wise correlations (which can be considered as a Gram-Matrix) is given in
\cite{li2017demystifying}. The authors treat style-transfer as a domain adaptation problem, and prove the equivalence between trying to match the Gram-Matrices of feature maps and minimizing the Maximum Mean Discrepancy (MMD) with the second-order polynomial kernel. 

One of the few attempts to apply neural-style transfer in speech synthesis is described in
\cite{chorowski2018using}. The authors employ a pretrained CNN and apply style-transfer to train a speech generation and voice conversion system. Moreover, they demonstrate the capacity of the intermediate tensors to encode speaker characteristics using the VCTK dataset. However, the scope of their work is different to ours, the network is not trained end-to-end while the representations are high-dimensional. Furthermore, their method is not tested on a speaker recognition benchmark containing noise and high intrinsic and extrinsic within-speaker session variability (such as VoxCeleb or NIST-SRE). Hence, the ability of their utterance-level representation to suppress session variability has not been examined.

\subsection{Pooling layers in speaker embeddings}
The vast majority of speaker embedding networks use the statistics pooling layer introduced by x-vectors, i.e. mean and standard deviation (std) for each frequency and channel combination \cite{snyder_interspeech_2017}. The main family of alternative pooling methods enhance these statistics with an attention mechanism. Examples are single and multi-head attention, as well as Net and Ghost-VLAD \cite{india2019self,safari2020self,nagrani2020voxceleb}. Both attention and VLAD methods employ a set of heads (i.e. trainable network parameters which are broadly analogous to the means of a mixture model) to cluster the output features into broad phonetic categories in an unsupervised and end-to-end fashion. The resulting statistics pooling is defined as the concatenation of the per-head statistics. Their main difference lies in the normalization; the attention normalizes only over the time axis while the VLAD over the attentive heads. Second order statistics (i.e. std features) may also be added as in the case of average pooling. More recent approaches exist, such as vector-based attentive pooling \cite{wu2020vector}, phonetically-aware attention \cite{zhou2019cnn}, and cross-attention \cite{kye2020cross}.

A second alternative to standard statistics pooling is the use of higher-order statistics (skewness and kurtosis) \cite{you2019multi}. Skewness is a measure of the asymmetry of a distribution with respect to its mode, while the kurtosis measures its “tailedness”. The authors did not obtain any consistent improvement compared to using merely first and second statistics, as in standard statistics pooling. Recently (and concurrently to our work) a set of alternative pooling methods were examined in \cite{wang2021revisiting}. The authors show that the use of second order statistics is more effective compared to first order, while they also perform experiments with using the full covariance matrix, without showing improvements over the standard statistics pooling. Their work is close to ours, although we differ in several ways, such as in the suggested pooling method and in the several implementation details (e.g. instead of covariances we propose frequency-dependent channel-wise correlation matrices).

\section{The proposed method}
\label{Sect:Proposed}
\subsection{Notation}
We denote by $\X = \{\x_t\}_{t=1}^{T_i}$ the sequence of acoustic features of an utterance with $T_i$ frames, which in our method are $F_i=80$ dimensional filterbank features. The architecture is a ResNet-34 and the output of the last ResNet block is a 3D tensor (we omit the batch axis) denoted by $\Y \in \R^{T\times F \times C}$, where $T$ denotes the size of the temporal axis, $F$ the size of the frequency axis, and $C$ the number of channels. Typical values for these quantities are $(T,F,C)=(50,10,256)$; note though that $T$ is variable and can be derived by $T_i$ divided by the cumulative temporal stride (equal to $2^3=8$ in our architecture). Similarly, $F=10$ since the cumulative frequency stride is again equal to 8, while $F_i=80$.

\subsection{Statistics pooling}

The statistics pooling layer in speaker embeddings networks with 2D CNN architectures is a concatenation of the mean and std of each of the $F\times C$ frequency-channel pairs $[\P^m;\P^s]$, where
\begin{equation}
\P^m_{f,c} = \frac{1}{T}\sum_{t}\Y_{t,f,c} 
\label{eq:spm}
\end{equation}
and
\begin{equation}
\P^s_{f,c} = \sqrt{\frac{1}{T}\sum_{t}(\Y_{t,f,c}-\P^m_{f,c})^2}  
\label{eq:sps}
\end{equation}

The speaker embeddings are extracted by projecting the (vectorized) pooling layer onto a lower dimensional space (e.g. $d_e = 256$) using a single linear layer.  

\subsection{Style modeling in images}
The style of an image in computer vision for a given convolutional layer is modeled by the symmetric positive-semidefinite matrix $\S$ with elements defined as
\begin{equation}
\S_{c,c'} =  \frac{1}{TF} \sum_{t,f} \Y_{t,f,c} \Y_{t,f,c'}
\label{eq:styleCV}
\end{equation}
Note though that images differ from spectral representations of acoustic signals. Image characteristics are considered invariant along both spatial axes, while spectral representations are invariant only along the time axis. In fact, this property explains the use of frequency-channel pairs in statistics pooling instead of channels alone, as shown in (\ref{eq:spm})-(\ref{eq:sps}). Due to this property of spectral representations, pooling along both time and frequency axes as in (\ref{eq:styleCV}) is expected to yield less discriminative speaker representations.  

\subsection{Pooling via frequency-dependent channel-wise correlations}
A natural modification of (\ref{eq:styleCV}) that addresses the above issue is to pool only along the time axis, i.e.
\begin{equation}
\S_{f,c,c'} =  \frac{1}{T} \sum_{t} \Y_{t,f,c} \Y_{t,f,c'}
\label{eq:xcorr}
\end{equation}
The 3D tensor $\S$ can be viewed as a list of $F$ symmetric and positive semidefinite matrices $\S_{f}=\S_{f,:,:}$ and has therefore $FC(C+1)/2$ free variables. As its size can be too large we propose to reduce it in a frequency-dependent manner as shown later. 

A further proposed modification is the use of mean and variance normalization along the time axis of $\Y$ for each frequency-channel pair, i.e. $\Z_{:,f,c} \xleftarrow{norm}\ \Y_{:,f,c}$. When variance normalization is applied, the variables in the diagonal of $\S_f$ (i.e. the variance of each channel-frequency combination) become equal to 1, and therefore they can be removed from the pooling layer resulting in $C(C-1)/2$ free variables for each frequency $f$ and $FC(C-1)/2$ overall. Moreover, $\S_{f,c,c'} \in [-1,1]$ for all $f,c,c'$. The pooling layer is followed by a linear layer which reduces the size of the representation to $d_e=256$ (the speaker embedding space), as in the standard statistics pooling.

Note that when mean and variance normalization are applied, the pooled representation encodes complementary information to that of the standard pooling; the means of each channel-frequency are being subtracted, while their variance is also set equal to one. 

\subsection{Additional extensions and modifications}
In the pooling method described above, we add the following operations in other to decrease the size of the pooling layer and regularize the network. Recall that regularization is needed in order to extract embeddings that generalize well to speakers unseen during training.

\subsubsection{Channel dropout}
Dropout is a standard choice for regularizing a network. We employ it in a channel-wise fashion, that is, we sample a 1D mask of size $C$ (with dropout probability $p_d=0.25$) and we broadcast it along the time and frequency axes. Channel-wise dropout is chosen because of its Bayesian interpretation as model averaging between models with variable number of convolutional filters \cite{gal2016theoretically}. 

\subsubsection{Frequency ranges}
As neighboring frequency bins in $\Y$ have similar statistics, we consider merging them into $F_r$ frequency ranges. After experimentation, we concluded that merging $f_r=2$ consecutive frequency bins in a non-overlapping manner yields more stable performance. Merging is implemented by reshaping $\Y$, so that $F_r \leftarrow Ff_r^{-1}$ and $T_r \leftarrow T f_r$. For example, when experimenting with $f_r=2$, we obtain $F_r=5$ and $T_r=100$ given that $F=10$ and $T=50$. Note that by setting $f_r=F$ we recover the original style-transfer used in computer vision, where pooling is performed along both spatial axes.

\subsubsection{Frequency-dependent channel reduction}
The number of channels in $\Y$ is $C=256$ which would result in $C(C-1)/2\approx 33K$ variables per frequency bin. So we transform $\Y$ as follows 
\begin{equation}
\Y_{t,f,c'} \leftarrow \sum_c \L_{f,c,c'}\Y_{t,f,c}. 
\label{eq.CRed}
\end{equation}
where $\L$ is a learnable tensor of shape $(F_r,C,C')$, acting on the channel axis and reducing its size from $C$ to $C'$ (we choose $C'=64$).
The rationale for using a 3D kernel instead of a 2D is that different linear combinations of channels should be better suited for obtaining speaker discriminative information from each frequency range.

\subsection{Summary and order of operations}
We summarise here the set of operations and their order we use in order to extract speaker embeddings. Recall that the output tensor of the last ResNet block is $\Y$ with shape $(T, F, C)$.
\begin{enumerate}
\item {\bf{Dropout}} We apply channel-wise dropout with $p_d$ by sampling a binary mask of shape $C$ and broadcasting it along the other two axes.   
\item {\bf{Frequency ranges}} We reshape the tensor $\Y$ with new shape $(T_r, F_r, C)$, by merging $f_r$ consecutive frequency bins into $F_r$ frequency ranges. 
\item {\bf{Channel Reduction}} We reduce the channel size by applying $\L$ as shown in (\ref{eq.CRed}), i.e. $\Y \xleftarrow{\L}\Y$, where the shape of $\Y$ becomes $(T_r, F_r, C')$ . 

\item {\bf{Normalization}} We apply mean and variance normalization along the time axis, i.e. $\Z \xleftarrow{norm}\Y$. 

\item {\bf{Pooling}} We apply pooling as shown in (\ref{eq:xcorr}), but using $\Z$ instead of $\Y$, by which we obtain $\S$, a 3D tensor with shape $(F_r, C',C')$. 

\item {\bf{Flattening}} We flatten the pooled representation to a vector of size $d_p = F_r C'(C'-1)/2$, which corresponds to the number of free variables in $\S$.  
\item {\bf Embedding} We reduce the size of the pooled representation vector to $d_e$ using a linear layer. The resulting vector is the speaker embedding.
\end{enumerate}

\section{Experiments}
\label{Sect:Exps}
\subsection{VoxCeleb 1 \& 2 datasets}
We evaluate the systems on the three VoxCeleb test sets, namely the original VoxCeleb 1 \emph{test} set (composed of only 40 speakers) \cite{Nagrani17} and the (much larger and representative) \emph{Extended} and \emph{Hard} sets  \cite{nagrani2020voxceleb}. 
As training set, we use a subset of the VoxCeleb 2 \emph{development} set which we augmented with babble, noise, reverberation, and music using Kaldi~\cite{povey2011kaldi,snyder2018x}. The training set contains 5750 speakers and approximately 4.9M utterances including the augmented versions. As a \emph{held-out} set we used 455 utterances (plus augmentations) from 90 training speakers.  

\subsection{Architecture, training, and loss function}
\label{ss:imp_det}
The backbone of the network is a 34-layer ResNet. All convolutional kernels are $3\times 3$, the strides are $(1,2,2,2)$ for the 4 blocks in both time and frequency axes, while the number of channels is $(64,128,256,256)$ (the first convolutional layer also outputs 64 channels). The number of convolutional layers per block is (3, 4, 6, 3). Finally, Squeeze and Excitation layers (with reduction ratio $r=4$) are added only to the first 2 blocks of the ResNet \cite{hu2018squeeze,desplanques2020ecapa}. 

The network is trained using multi-speaker classification. As optimization criterion we employ the Additive Angular Margin (AAM) loss, which has shown notable improvements over the plain Cross Entropy loss \cite{deng2019arcface}. We set the scale coefficient of the loss equal to 30, while for the margin coefficient we adopt a curriculum learning approach and progressively increase it as (0.1,0.2,0.3) during training \cite{chung2020defence}.  

As optimizer we use stochastic gradient descent with momentum equal to 0.9. The minibatch size is 256, however to train the model with a single GPU we split the minibatch into 16 ``microbatches'' of 16 examples each and use gradient accumulation. The initial learning rate (LR) is equal to 0.2 which we divide by 2 when the loss does not improve for more than 3000 model updates in the held-out set (the final LR is 0.2/64). The experiments are conducted using TensorFlow \cite{zeinali2019improve}.

We emphasize that the model architecture, the training method, and its hyperparameters are optimized with the baseline, where mean and std features are used for pooling. Moreover, we did not observe any differences in time required per epoch between the baseline and the proposed architectures. 

\begin{table*}[th]
  \caption{
  Results on the Original, Hard, and Extended VoxCeleb test sets.
  } \vspace*{-2.5mm}
  \label{tab:results}
  \centering
   \begin{tabular}{ c l c c c c c c  c c  c c }

    \toprule
     & \textbf{System} & & & & &  \multicolumn{2}{c}{\textbf{ VoxCeleb-O }} & \multicolumn{2}{c}{\textbf{VoxCeleb-H}} & \multicolumn{2}{c}{\textbf{VoxCeleb-E}} \\
     & & & & & & minDCF & EER(\%) & minDCF & EER(\%) & minDCF & EER(\%)\\
    \midrule
    B1 & Baseline & \multicolumn{4}{c}{(mean \& std pooling)}  & 0.091 & 1.40 & 0.145 & 2.48 & 0.090 & 1.43 \\
    
    B2 & Baseline & \multicolumn{4}{c}{(std pooling)}  & 0.104 & 1.56 & 0.154 & 2.61 & 0.099 & 1.54 \\
    
    \midrule
       &     & $C'$ & $F_r$ & $\L$ & Normalization & & & & & & \\
    \midrule
    
    P1 & Proposed & 128 & 1 & 2D & mean &  0.126 & 1.74 & 0.177 & 2.93 & 0.114 & 1.74 \\
    P2 & Proposed & 64 & 5 & 2D & mean  & 0.106 & 1.66 & 0.166 & 2.80 & 0.104 & 1.67 \\
    P3 & Proposed & 64 & 5 & 3D & mean   & 0.089 & 1.55 & 0.148 & 2.49 & 0.092 & 1.48 \\ 
    P4 & Proposed & 64 & 10 & 3D & mean   & 0.109 & 1.70 & 0.159 & 2.70 & 0.103 & 1.62 \\  
    P5 & Proposed & 128 & 1 & 2D & mean \& var. & 0.102 & 1.56 & 0.158 & 2.63 & 0.101 & 1.57 \\ 
    P6 & Proposed & 64 & 5 & 2D & mean \& var. & 0.095 & 1.46 & 0.143 & 2.40 & 0.091 & 1.41 \\
    P7 & Proposed & 64 & 5 & 3D & mean \& var. & \bf{0.071} & \bf{1.16} & \bf{0.128} & \bf{2.17} & \bf{0.079} & \bf{1.22} \\
    P8 & Proposed & 64 & 10 & 3D & mean \& var. & 0.085 & 1.41 & 0.137 & 2.33 & 0.087 & 1.39 \\ 
    \midrule
    B1s & \multicolumn{5}{c}{B1 with adaptive score normalization (as-norm)} & 0.079 & 1.26 & 0.127 & 2.23 & 0.080 & 1.31 \\
    P7s & \multicolumn{5}{c}{P7 with adaptive score normalization (as-norm)} & 0.059 & 1.07 & 0.115 & 1.99 & 0.071 & 1.13 \\
    \bottomrule
    \vspace*{-4mm}

  \end{tabular}
\end{table*}

\subsection{Experimental Results}
To evaluate our method we experiment we examine several configurations and ablations. The results we obtained are given in Table \ref{tab:results}, and are obtained using cosine-similarity, without score or any other normalization. For the two baseline experiments we train and evaluate a ResNet-34 with the standard temporal pooling. The first includes mean and std features while the second only std (suggested in \cite{wang2021revisiting}). We emphasize that for the baseline experiments the tensor operations are not included, i.e. mean and std pooling is performed directly on the output of the last ResNet block.    

The first experiment using a variant of the proposed pooling (P1) is essentially a direct implementation of the pooling as suggested in image style-transfer (recall that pooling with $F_r=1$ is equivalent to (\ref{eq:styleCV}), i.e. to pooling along both time and frequency axes.). Two such networks are examined, one with mean and a second one with mean and variance normalization, denoted by P1 and P5, respectively. Note that when mean-only normalization is applied, the number of free-parameters is $F_rC_r(C_r+1)/2$, i.e. the size of the pooled representation vector is increased by $F_rC_r$. By comparing P1 and P5 we observe that variance normalization helps significantly the model to attain results that are very close to the baseline B1. We should also note that for the experiments with $F_r=1$, we set $C'=128$ in order to have pooling layers of same order of magnitude, since their size grows linearly with $F_r$.

In the next set of experiments we use either $F_r=5$ or $F_r=10$, where the latter indicates no reshaping of the tensor $\Y$ (as it corresponds to $f_r=1$). The results confirm that the frequency axis should be treated differently compared to the time axis, and pooling should result in a 3D tensor, as shown in (\ref{eq:xcorr}). They moreover demonstrate that reshaping the tensor by a factor of $f_r=2$ (i.e. $F_r=5$) yields the best performance.

We may also observe that the suggested frequency-dependent channel reduction (i.e. using a 3D tensor $\L$ instead of a 2D) yields consistently improved performance (by comparing P2 with P3, and P6 with P7). Allowing each frequency range to have its own weighted combination of channels increases the expressiveness of the pooled representation. We also mention that we attempted to combine our best system (P7) with B1 in the pooling layer. The improvement in performance attained was small. For example the EER dropped by 0.03\% and the minDCF by 0.002 in both in VoxCeleb-E and VoxCeleb-H. Although further experimentation is required, it suggests that combining the correlations-based with mean and std features does not yield any notable improvement. We also compare B1 with P7 using score normalization (as-norm). We observe that the gains attained by the suggested pooling (P7s) over mean-std pooling (B1s) remain even when as-norm is applied.   

Finally, all experiments underline the importance of mean and variance normalization (P5-8), as opposed to mean-only (P1-4). In other words, transforming the Gram-Matrices to correlation matrices results in more speaker-discriminative representations. This finding challenges the common view of frame aggregation being the natural means of extracting speaker-relevant information from utterances. Correlations, as {\it intrinsically normalized} statistical quantities, should be more resilient as features to the diverse factors of within-speaker variability. Temporal aggregation may not be the most effective way to remove such undesired sources of variability. Removing them via modeling correlations between different representations of the utterances (such as time-channel maps of given frequency ranges) may attain this goal more effectively and naturally.

\section{Conclusions and research directions}
\label{Sect:Conc}
In this paper, we introduced a new way of pooling information extracted from a CNN for modeling speaker information. The approach is inspired by neural style-transfer in computer vision, where style is modeled by a Gram-Matrix of channel-wise dot-products. We proposed a way to adapt this method to speech signals, e.g. by calculating several such Gram-Matrices, one per frequency range. We further demonstrated that transforming them to correlation matrices increases the speaker-discriminability of the extracted representations. Our experiments on VoxCeleb indicate that the proposed method yields notable improvements over the standard statistics pooling.  

There are several ways by which our approach can be extended. For estimating correlations between channels, the dot-product can be replaced by kernel functions, some of which have been examined in image style-transfer \cite{jing2019neural}. The property of Gram-Matrix being a symmetric positive semi-definite matrix can be explored in several ways, using e.g. statistical divergences (e.g. Kullback-Leibler, Jensen-Shannon, a.o.). Alternative training/finetuning schemes (e.g. prototypical, triplet, contrastive and self-supervised losses \cite{chung2020defence,zhang2017end,ravanelli2019learning,stafylakis2019self}) that do not require a classification head are probably better suited for such kind of experimentation. Finally, the whole chain of transforms we apply on the tensor $\Y$ is not a result of exhaustive experimentation, leaving room for further improvements.

\section{Acknowledgements}
The work was supported by Czech National Science Foundation (GACR) project ``NEUREM'' No. 19-26934X, and European Union’s Horizon 2020 project No. 833635 ROXANNE. This project has received funding from the European Union’s Horizon 2020 research and innovation programme under grant agreement No 101007666 / ESPERANTO / H2020-MSCA-RISE-2020. The authors do not see any significant ethical or privacy concerns that would prevent the processing of the data used in the study. The datasets do contain personal data, and these are processed in compliance with the GDPR and national law.

\bibliographystyle{IEEEtran}

\bibliography{mybib}


\end{document}